\documentclass{elsart}

\usepackage{amsmath,amssymb,graphicx,subfigure,url}

\journal{JSTAT}

\begin{document}

% FRONT MATTER ----------------------------------------------------------------------------------------------------------------------------------------

\begin{frontmatter}

\title{A $\boldsymbol{\kappa}$-generalized statistical mechanics approach to income analysis}

\author[Ancona]{F. Clementi\corauthref{cor}},
\corauth[cor]{Corresponding author: Tel.: +39--071--22--07--103; fax: +39--071--22--07--102.}
\ead{fabio.clementi@univpm.it}
\author[Ancona]{M. Gallegati},
\ead{mauro.gallegati@univpm.it}
\author[Torino]{G. Kaniadakis}
\ead{giorgio.kaniadakis@polito.it}

\address[Ancona]{Department of Economics, Polytechnic University of Marche, Piazzale R. Martelli 8, 60121 Ancona, Italy}
\address[Torino]{Department of Physics, Polytechnic University of Turin, Corso Duca degli Abruzzi 24, 10129 Torino, Italy}

\begin{abstract}
This paper proposes a statistical mechanics approach to the analysis of income distribution and inequality. A new distribution function, having its roots in the framework of $\kappa$-generalized statistics, is derived that is particularly suitable to describe the whole spectrum of incomes, from the low-middle income region up to the high-income Pareto power-law regime. Analytical expressions for the shape, moments and some other basic statistical properties are given. Furthermore, several well-known econometric tools for measuring inequality, which all exist in a closed form, are considered. A method for parameter estimation is also discussed. The model is shown to fit remarkably well the data on personal income for the United States, and the analysis of inequality performed in terms of its parameters reveals very powerful.
\end{abstract}

\begin{keyword}
Personal income distribution\sep inequality\sep $\kappa$-generalized statistics
\PACS 02.50.Ng\sep 02.60.Ed\sep 89.65.Gh
\end{keyword}

\end{frontmatter}

% INTRODUCTION ----------------------------------------------------------------------------------------------------------------------------------------

\section{Introduction}

Measurement of income inequality to evaluate social welfare is of particular interest to economics. Since the size distribution of income is the basis of inequality measures, correct specification of the income density function is of great importance. The study of the income size distribution has a long history. Pareto \cite{Pareto1897} apparently was responsible for the first attempt at defining a general ``law'' that tried to explain the regularities of observed distributions. Let $P_{\geq}\left(x\right)$ be the percentage of individuals with incomes greater than or equal to $x$. Then, the (strong) Pareto law asserts that
\begin{equation}
P_{\geq}\left(x\right)=
\begin{cases}
\left(x/x_{0}\right)^{-\alpha}&\text{when}\quad x_{0}\leq x<\infty\\
1&\text{when}\quad x<x_{0}
\end{cases},
\label{eq:Equation1}
\end{equation}
for some $x_{0},\alpha>0$ and the support of $P_{\geq}\left(x\right)$ is $\left[x_{0},+\infty\right)$.

Available empirical work leaves little doubt that Pareto law, as it stands, does not account satisfactorily for a wide range of incomes. Subsequently, the use of other density functions to model the income distribution, such as the lognormal \cite{AitchisonBrown} or gamma \cite{SalemMount1974}, has been advocated. However, rapidly accruing evidence showed that the lognormal and gamma distributions fit the data relatively well in the middle range of income but tend to exaggerate the skewness and perform poorly in the upper end \cite{McDonaldEtAl}. Furthermore, if one's attention is restricted to the upper tail of the distributions, the evidence does not contradict the (strong) Pareto law, provided that the chosen $x_{0}$ is large enough. This suggests that observed distributions obey a weak version of the Pareto law \cite{Mandelbrot1960}, i.e.
\begin{equation}
\lim_{x\rightarrow\infty}\frac{P_{\geq}\left(x\right)}{\left(x/x_{0}\right)^{-\alpha}}=1
\label{eq:Equation2}
\end{equation}
for $P_{\geq}\left(x\right)$ with support $\left[a,+\infty\right)$ and $a\geq0$, and some well-known density functions that have been proposed and implemented in the literature asymptotically approach (rather than coincide with) the Pareto distribution. Among these, the Singh-Maddala \cite{SinghMaddala1976} and Dagum \cite{Dagum1977} distributions have shown them to be a good compromise between parsimony and goodness-of-fit in many instances.

Distributions exhibiting Pareto fat tails have been observed experimentally also in physical statistical systems. Since they differs from the ordinary exponential distributions, this fact needs a theoretical explanation. In the last few decades several physical mechanisms have been considered in order to justify the non-exponential equilibrium distributions. For instance, deviations from the exponential distribution can be originated by quantum
effects \cite{KaniadakisLavagnoQuarati} or by anomalous diffusion which introduces nonlinearities in the particle kinetics both in the Fokker-Planck \cite{KaniadakisLapentaQuarati} and in the Boltzmann picture \cite{BiroKaniadakis2006} of the system.

In physics, the deviation of the distribution function from the exponential distribution, i.e. the power-law tails, presents at high energies. Then the relativistic origin of this effect appears as the more natural. Recently, a statistical distribution based on the following one-parameter deformation of the exponential function
\begin{equation}
\exp_{\kappa}\left(x\right)=\left(\sqrt{1+\kappa^{2}x^{2}}+\kappa x\right)^{1/\kappa},
\label{eq:Equation3}
\end{equation}
with $x\in\mathbf{R}$ and $\kappa\in\left[0,1\right)$, has been proposed by one the authors \cite{Kaniadakis2001}. The $\kappa$-exponential can be inverted easily and the $\kappa$-logarithm is defined by
\begin{equation}
\ln_{\kappa}\left(x\right)=\frac{x^{\kappa}-x^{-\kappa}}{2\kappa},
\label{eq:Equation4}
\end{equation}
with $x>0$ and $\kappa\in\left[0,1\right)$.

The mechanism generating the latter deformation is originated by the microscopic Einstein relativistic dynamics \cite{Kaniadakis} and for the deformation parameter it results $\kappa\propto1/c$, being $c$ the light speed. The value of $\kappa\neq0$ is due to the finite value of the light speed and the deformation is originated ultimately by the Lorentz transformations.

In order to better explain how the special relativity conditioned the form of the $\kappa$-exponential function we recall that the relativistic momenta $x$ and $y$ of two identical particles $A$ and $B$ which move in the same direction, if observed in the rest frame of the particle $B$ becomes $x^{'}=x\stackrel{\kappa}{\oplus}\left(-y\right)$ and $y^{'}=0$ respectively. The relativistic composition law $\stackrel{\kappa}{\oplus}$ for the dimensionless momenta, according to the Lorentz transformations, is a generalized sum defined through
\begin{equation}
x\stackrel{\kappa}{\oplus}y=x\sqrt{1+\kappa^{2}y^{2}}+y\sqrt{1+\kappa^{2}x^{2}}.
\label{F1}
\end{equation}
The $\kappa$-exponential satisfies the functional equation
\begin{equation}
\exp_{\kappa}\left(x\stackrel{\kappa}{\oplus}y\right)=\exp_{\kappa}\left(x\right)\exp_{\kappa}\left(y\right),
\label{F2}
\end{equation}
which, in the classical limit $\kappa\rightarrow0$, where $\exp_{\kappa}\left(x\right)\rightarrow\exp\left(x\right)$ and $x\stackrel{\kappa}{\oplus}y \rightarrow x+y$, reduces to the classical equation $\exp(x+y)=\exp(x)\exp(y)$ of the ordinary exponential function.

The relativistic sum defined in Equation \eqref{F1} induces a relativistic generalized mathematics where all the mathematical operators and functions emerge properly deformed. For instance the ordinary derivative operator transforms into the $\kappa$-derivative given by
\begin{equation}
\frac{\mathrm{d}}{\mathrm{d}_{\kappa}x}=\sqrt{1+\kappa^{2}x^{2}}\frac{\mathrm{d}}{\mathrm{d}x}.
\label{F3}
\end{equation}
Within this theoretical framework the $\kappa$-exponential emerges as the relativistic generalization of the ordinary exponential. In particular it holds the relationship
\begin{equation}
\frac{\mathrm{d}}{\mathrm{d}_{\kappa}x}\exp_{\kappa}\left(x\right)=\exp_{\kappa}\left(x\right),
\label{F4}
\end{equation}
which is the relativistic generalization of the classical equation $\left(\mathrm{d}/\mathrm{d}x\right)\exp\left(x\right)=\exp\left(x\right)$ involving the ordinary derivative and exponential.

\sloppy The ordinary exponential function $\exp(x)$ emerges both at low energies, being
\begin{equation}
\exp_{\kappa}\left(x\right){\atop\stackrel{\textstyle\sim}{\scriptstyle x\rightarrow0}}\exp\left(x\right),
\label{eq:Equation5}
\end{equation}
as well as when the deformation parameter $\kappa$ approaches zero, i.e. $\lim_{\kappa\rightarrow0}\exp_{\kappa}\left(x\right)=\exp\left(x\right)$. On the contrary, for high values of $x$ the function $\exp_{\kappa}(x)$ presents power-law tails
\begin{equation}
\exp_{\kappa}\left(x\right){\atop\stackrel{\textstyle\sim}{\scriptstyle x\rightarrow\pm\infty}}\big|2\kappa x\big|^{\pm1/\left|\kappa\right|}.
\label{eq:Equation6}
\end{equation}

\fussy The statistical mechanics based on $\exp_{\kappa}\left(x\right)$ preserves the Legendre structures of the ordinary statistical mechanics and the underlying entropy is stable \cite{AbeKaniadakisScarfone}. The relevant statistical distribution at low energies is just the Boltzmann distribution according to Equation \eqref{eq:Equation5}, while at high energies presents presents power-law tails according to Equation \eqref{eq:Equation6}.

The particularly interesting mathematical properties of the $\kappa$-exponential permit us to see this function as a very flexible mathematical tool in order to study efficiently also non-physical systems. Indeed, in the literature this function have been used extensively in several fields beyond the relativity, e.g. in dynamical systems at the edge of chaos, in fractal systems, in game theory, in error theory, in economics and so on.

On the other hand, it is well known that the Einstein relativity has the same basis of the Galilei relativity of classical physics, except for the presence of an extra Einstein principle, asserting that the information propagates with a finite speed ($\kappa\neq0$) and not instantaneously ($\kappa=0$) as professed in classical physics. This so natural relativistic principle relegates the ordinary exponential at the status of an abstract and nonphysical function and legitimates the use of the function $\kappa$-exponential in the analysis of real systems.

In this paper we exploit the deformed exponential function as a functional relationship that is more flexible than the standard one to build statistical models by adapting it to the context of income size distribution. Using such a deformed exponential function is attractive because it allows one to statistically describe the whole spectrum of the size distribution of incomes, ranging from the low region to the middle region, and up to the Pareto tail. The $\kappa$-deformed statistical model leads to a more general formulation that contains both Pareto and stretched exponential distributions as limiting cases.

The rest of the paper is organized as follows. In Section \ref{sec:TheKappaGeneralizedStatisticalDistribution}, we examine the theoretical properties of what we refer to as the $\kappa$-generalized distribution and show how it is able to account for some basic stylized facts of personal income data, such as the weak Pareto law and possessing at least one interior mode. In Section \ref{sec:EmpiricalApplicationToUSIncomeData}, in order to test the performance of the proposed distribution, we provide an empirical application to the U.S. personal income data. The paper is concluded in Section \ref{sec:ConcludingRemarks}.

% THE k-GENERALIZED STATISTICAL DISTRIBUTION ----------------------------------------------------------------------------------------------------------

\section{The $\boldsymbol{\kappa}$-generalized statistical distribution}
\label{sec:TheKappaGeneralizedStatisticalDistribution}

In view of their importance for the proposed statistical model, in the following we firstly recall some basic mathematical properties of the $\kappa$-deformed exponential and logarithm functions. Then we give formulas for the shape, moments and standard tools for inequality measurement. These include, among others, the ubiquitous Lorenz curve and the associated Gini measure of income inequality. In addition, we also discuss a method for parameter estimation.

% The k-deformed exponential and logarithm functions --------------------------------------------------------------------------------------------------

\subsection{The $\kappa$-deformed exponential and logarithm functions}

The power-law asymptotic behavior of $\exp_{\kappa}\left(x\right)$ as given by Equation \eqref{eq:Equation6} reappears also in the function $\ln_{\kappa}\left(x\right)$, namely
\begin{subequations}
\begin{equation}
\ln_{\kappa}\left(x\right){\atop\stackrel{\textstyle\sim}{\scriptstyle x\rightarrow0^+}}-\frac{1}{2\left|\kappa\right|}x^{-\left|\kappa\right|}
\label{eq:Equation7a}
\end{equation}
and
\begin{equation}
\ln_{\kappa}\left(x\right){\atop\stackrel{\textstyle\sim}{\scriptstyle x\rightarrow+\infty}}\frac{1}{2\left|\kappa\right|}x^{\left|\kappa\right|}.
\label{eq:Equation7b}
\end{equation}
\end{subequations}

Like the ordinary functions, also the deformed ones have the properties
\begin{subequations}
\begin{equation}
\exp_{\kappa}\left(x\right)\exp_{\kappa}\left(-x\right)=1,
\label{eq:Equation8a}
\end{equation}
\begin{equation}
\ln_{\kappa}\left(1/x\right)=-\ln_{\kappa}\left(x\right)
\label{eq:Equation8b}
\end{equation}
\end{subequations}
and
\begin{subequations}
\begin{equation}
\left[\exp_{\kappa}\left(x\right)\right]^{r}=\exp_{\kappa/r}\left(rx\right),
\label{eq:Equation9a}
\end{equation}
\begin{equation}
\ln_{\kappa}\left(x^{r}\right)=r\ln_{r\kappa}\left(x\right).
\label{eq:Equation9b}
\end{equation}
\end{subequations}

The Taylor expansions of the functions $\exp_{\kappa}\left(x\right)$ and $\ln_{\kappa}\left(x\right)$ are given by
\begin{subequations}
\begin{equation}
\exp_{\kappa}\left(x\right)=1+x+\frac{x^{2}}{2}+\left(1-\kappa^{2}\right)\frac{x^{3}}{3!}+\ldots
\label{eq:Equation10a}
\end{equation}
and
\begin{equation}
\ln_{\kappa}\left(1+x\right)=x-\frac{x^{2}}{2}+\left(1+\frac{\kappa^{2}}{2}\right)\frac{x^{3}}{3}-\ldots,
\label{eq:Equation10b}
\end{equation}
\end{subequations}
respectively, and hold for $x\rightarrow0$.

% The distribution and its properties -----------------------------------------------------------------------------------------------------------------

\subsection{The distribution and its properties}

In the last few years the $\kappa$-exponential function was adopted successfully to analyze also non-physical systems, including economic systems. In particular, the $\kappa$-deformation has been employed in order to propose the so-called $K$-deformed multinomial logit model to study differentiated product markets \cite{BolducJayetRajaoarison} and to model the personal income distribution \cite{ClementiDiMatteoGallegatiKaniadakis}. In this latter application the distribution function was defined through
\begin{equation}
P_{\geq}\left(x\right)=\exp_{\kappa}\left(-\beta x^{\alpha}\right),
\label{eq:Equation11}
\end{equation}
where $x\in\mathbf{R}$, $\alpha,\beta>0$ and $\kappa\in[0,1)$. The income variable $x$ is defined as $x=z/\left\langle z\right\rangle$, being $z$ the absolute personal income and $\left\langle z\right\rangle$ its mean value. The corresponding density reads
\begin{equation}
p\left(x\right)=\frac{\alpha\beta x^{\alpha-1}\exp_{\kappa}\left(-\beta x^{\alpha}\right)}{\sqrt{1+\kappa^2\beta^2x^{2\alpha}}},
\label{eq:Equation12}
\end{equation}
while
the quantile function is available in the following closed form
\begin{equation}
x\left(u\right)=\beta^{-1/\alpha}\left[-\ln_{\kappa}\left(1-u\right)\right]^{1/\alpha},
\label{eq:Equation13}
\end{equation}
with $u=P_{<}\left(x\right)=1-P_{\geq}\left(x\right)$ and $0\leq u\leq1$.

As $\kappa\rightarrow0$ this model tends to the stretched exponential distribution; it can be easily verified that
\begin{subequations}
\begin{equation}
\lim_{\kappa\rightarrow0}P_{\geq}\left(x\right)=\exp\left(-\beta x^{\alpha}\right)
\label{eq:Equation14a}
\end{equation}
and
\begin{equation}
\lim_{\kappa\rightarrow0}p\left(x\right)=\alpha\beta x^{\alpha-1}\exp\left(-\beta x^{\alpha}\right).
\label{eq:Equation14b}
\end{equation}
\end{subequations}

For low incomes ($x\rightarrow0$) the distribution behaves similarly to the stretched exponential Equation \eqref{eq:Equation14a} and Equation \eqref{eq:Equation14b}, while at high incomes it approaches a Pareto distribution with scale $\left(2\beta\kappa\right)^{-1/\alpha}$ and shape $\alpha/\kappa$, i.e.
\begin{subequations}
\begin{equation}
P_{\geq}\left(x\right){\atop\stackrel{\textstyle\sim}{\scriptstyle x\rightarrow+\infty}}\left(2\beta\kappa\right)^{-1/\kappa}x^{-\alpha/\kappa}
\label{eq:Equation15}
\end{equation}
and
\begin{equation}
p\left(x\right){\atop\stackrel{\textstyle\sim}{\scriptstyle x\rightarrow+\infty}}\frac{\alpha}{\kappa}\left(2\beta\kappa\right)^{-1/\kappa}x^{-\left(\frac{\alpha}{\kappa}+1\right)},
\label{eq:Equation16}
\end{equation}
\end{subequations}
thus satisfying the weak Pareto law \cite{Kakwani1980}
\begin{equation}
\lim_{x\to\infty}\frac{xp\left(x\right)}{P_{\geq}\left(x\right)}=\frac{\alpha}{\kappa},
\label{eq:Equation17}
\end{equation}
which is a rephrased version of Equation \eqref{eq:Equation2}.

From Equation \eqref{eq:Equation13} we easily determine that the median of the distribution is
\begin{equation}
x_{\mathrm{med}}=\beta^{-1/\alpha}\left[\ln_{\kappa}\left(2\right)\right]^{\frac{1}{\alpha}}.
\label{eq:Equation18}
\end{equation}

The mode is at
\begin{equation}
\begin{split}
x_{\mathrm{mode}}=&\beta^{-1/\alpha}\Biggl\{\left[\frac{\alpha^{2}+2\kappa^{2}\left(\alpha-1\right)}{2\kappa^{2}\left(\alpha^{2}-\kappa^{2}\right)}\right]\Biggr.\\
&\left.\cdot\left(\sqrt{1+\frac{4\kappa^{2}\left(\alpha^{2}-\kappa^{2}\right)\left(\alpha-1\right)^{2}}{\left[\alpha^{2}+2\kappa^{2}\left(\alpha-1\right)\right]^{2}}}-1\right)\right\}^{\frac{1}{2\alpha}}
\end{split}
\label{eq:Equation19}
\end{equation}
if $\alpha>1$; otherwise, the distribution is zero-modal with a pole at the origin.

% Moments and other basic properties ------------------------------------------------------------------------------------------------------------------

\subsection{Moments and other basic properties}

\sloppy The moment about zero of order $r-1$ of $\exp_{\kappa}\left(-\beta x^{\alpha}\right)$, with $0<r<1/\kappa$, can be obtained in closed form and is given by
\begin{equation}
\int\limits_{0}^{\infty}x^{r-1}P_{\geq}\left(x\right)\operatorname{d} x=\frac{1}{\alpha}\frac{\left(2\beta\kappa\right)^{-\frac{r}{\alpha}}}{1+\frac{r}{\alpha}\kappa}\frac{\Gamma\left(\frac{1}{2\kappa}-\frac{r}{2\alpha}\right)}{\Gamma\left(\frac{1}{2\kappa}+\frac{r}{2\alpha}\right)}\Gamma\left(\frac{r}{\alpha}\right),
\label{eq:Equation20}
\end{equation}
where $\Gamma\left(\cdot\right)$ denotes the gamma function. Therefore, the moment of order $r$ expressed in terms of the density function Equation \eqref{eq:Equation12}, i.e. $\mu^{'}_{r}=r\int_{0}^{\infty}x^{r-1}P_{\geq}\left(x\right)\operatorname{d} x=\int_{0}^{\infty}x^{r}p\left(x\right)\operatorname{d} x$,  equals
\begin{equation}
\mu^{'}_{r}=\frac{r}{\alpha}\frac{\left(2\beta\kappa\right)^{-\frac{r}{\alpha}}}{1+\frac{r}{\alpha}\kappa}\frac{\Gamma\left(\frac{1}{2\kappa}-\frac{r}{2\alpha}\right)}{\Gamma\left(\frac{1}{2\kappa}+\frac{r}{2\alpha}\right)}\Gamma\left(\frac{r}{\alpha}\right).
\label{eq:Equation21}
\end{equation}
Specifically, $\mu^{'}_{1}=m$ is the mean of the distribution and the variance, $\sigma^{2}=\mu^{'}_{2}-m^{2}$, is defined as
\begin{equation}
\sigma^{2}=\left(2\beta\kappa\right)^{-\frac{2}{\alpha}}\left\{\frac{\Gamma\left(1+\frac{2}{\alpha}\right)}{1+2\frac{\kappa}{\alpha}}\frac{\Gamma\left(\frac{1}{2\kappa}-\frac{1}{\alpha}\right)}{\Gamma\left(\frac{1}{2\kappa}+\frac{1}{\alpha}\right)}-\left[\frac{\Gamma\left(1+\frac{1}{\alpha}\right)}{1+\frac{\kappa}{\alpha}}\frac{\Gamma\left(\frac{1}{2\kappa}-\frac{1}{2\alpha}\right)}{\Gamma\left(\frac{1}{2\kappa}+\frac{1}{2\alpha}\right)}\right]^{2}\right\}.
\label{eq:Equation22}
\end{equation}
Hence, the coefficient of variation, $CV_{\kappa}=\sigma/m$, equals
\begin{equation}
CV_{\kappa}=\sqrt{2\frac{\left(\alpha+\kappa\right)^{2}}{\alpha+2\kappa}\frac{\Gamma\left(\frac{2}{\alpha}\right)}{\Gamma^{2}\left(\frac{1}{\alpha}\right)}\frac{\Gamma\left(\frac{1}{2\kappa}-\frac{1}{\alpha}\right)}{\Gamma\left(\frac{1}{2\kappa}+\frac{1}{\alpha}\right)}\frac{\Gamma^{2}\left(\frac{1}{2\kappa}+\frac{1}{2\alpha}\right)}{\Gamma^{2}\left(\frac{1}{2\kappa}-\frac{1}{2\alpha}\right)}-1}.
\label{eq:Equation23}
\end{equation}

\fussy It is also possible to define the standardized measures $\gamma_{1}=\mu_{3}/\sigma^{3}$ and $\gamma_{2}=\mu_{4}/\sigma^{4}$ of skewness and kurtosis, respectively, given by
\begin{equation}
\gamma_{1}=\frac{\mu^{'}_{3}-3\mu^{'}_{2}m+2m^{3}}{\sigma^{3}}
\label{eq:Equation24}
\end{equation}
and
\begin{equation}
\gamma_{2}=\frac{\mu^{'}_{4}-4\mu^{'}_{3}m-6\mu^{'}_{2}m^{2}-3m^{4}}{\sigma^{4}},
\label{eq:Equation25}
\end{equation}
where
\begin{equation}
\mu_{r}=\sum^{r}_{j=0}{r \choose j}\left(-1\right)^{r-j}\mu^{'}_{j}m^{r-j}
\label{eq:Equation26}
\end{equation}
is the moment about the mean of order $r$.

% Lorenz curve and inequality measures ----------------------------------------------------------------------------------------------------------------

\subsection{Lorenz curve and inequality measures}

For a discussion of income inequality, the standard practice adopts the concept of concentration of incomes as defined by Lorenz \cite{Lorenz1905}. The so-called Lorenz curve measures the cumulative fraction of population with incomes below $x$ along the horizontal axis, and the fraction of the total income this population accounts for along the vertical axis. The points plotted for the various values of $x$ trace out a curve below the $45^{\circ}$ line sloping upwards to the right from the origin.

In statistical terms, for any general distribution supported on the nonnegative half-line with a finite and positive first moment the Lorenz curve is available in terms of the first-moment distribution $L\left(u\right)=m^{-1}\int^{x}_{0}x^{'}p\left(x^{'}\right)\operatorname{d} x^{'}$. Thus we have the Lorenz curve for the $\kappa$-generalized distribution as follows
\begin{equation}
\begin{split}
L_{\kappa}\left(u\right)=&1-\frac{1+\frac{\kappa}{\alpha}}{2\Gamma\left(\frac{1}{\alpha}\right)}\frac{\Gamma\left(\frac{1}{2\kappa}+\frac{1}{2\alpha}\right)}{\Gamma\left(\frac{1}{2\kappa}-\frac{1}{2\alpha}\right)}\left\{2\alpha\left(2\kappa\right)^{\frac{1}{\alpha}}\left(1-u\right)\left[\ln_{\kappa}\left(\frac{1}{1-u}\right)\right]^{\frac{1}{\alpha}}\right.\\
&\Biggl.+B_{X}\left(\frac{1}{2\kappa}-\frac{1}{2\alpha},\frac{1}{\alpha}\right)+B_{X}\left(\frac{1}{2\kappa}-\frac{1}{2\alpha}+1,\frac{1}{\alpha}\right)\Biggr\},
\end{split}
\label{eq:Equation27}
\end{equation}
where $B_{X}\left(\cdot,\cdot\right)$ is the incomplete beta function with $X=\left(1-u\right)^{2\kappa}$.

The related Gini coefficient of inequality \cite{Gini1914} can be easily derived using its representation in terms of order statistics \cite{ArnoldLaguna1977}, i.e. $G=1-m^{-1}\int^{\infty}_{0}\left[P_{\geq}\left(x\right)\right]^{2}\operatorname{d} x$; this yields
\begin{equation}
G_{\kappa}=1-\frac{2\alpha+2\kappa}{2\alpha+\kappa}\frac{\Gamma\left(\frac{1}{\kappa}-\frac{1}{2\alpha}\right)}{\Gamma\left(\frac{1}{\kappa}+\frac{1}{2\alpha}\right)}\frac{\Gamma\left(\frac{1}{2\kappa}+\frac{1}{2\alpha}\right)}{\Gamma\left(\frac{1}{2\kappa}-\frac{1}{2\alpha}\right)}.
\label{eq:Equation28}
\end{equation}

Furthermore, other summary inequality measures can be derived which are well-known and of widespread use in the econometric literature. For instance, in the context of the $\kappa$-deformed distribution the generalized entropy (GE) class of inequality measures \cite{CowellEtAl} assumes the form
\begin{equation}
GE_{\kappa}\left(\theta\right)=\frac{1}{\theta^{2}-\theta}\left\{m^{-\theta}\left[\frac{\left(2\beta\kappa\right)^{-\frac{\theta}{\alpha}}}{1+\frac{\theta}{\alpha}\kappa}\frac{\Gamma\left(\frac{1}{2\kappa}-\frac{\theta}{2\alpha}\right)}{\Gamma\left(\frac{1}{2\kappa}+\frac{\theta}{2\alpha}\right)}\Gamma\left(1+\frac{\theta}{\alpha}\right)\right]-1\right\},
\label{eq:Equation29}
\end{equation}
with $\theta\neq0,1$. Equation \eqref{eq:Equation29} defines a class because the index $GE_{\kappa}\left(\theta\right)$ assumes different forms depending on the value assigned to $\theta$. From an operational point of view, two limiting cases of Equation \eqref{eq:Equation29} are of particular interest for inequality measurement: the mean logarithmic deviation index, $MLD_{\kappa}=\lim_{\theta\rightarrow0}GE_{\kappa}\left(\theta\right)$, given by
\begin{equation}
MLD_{\kappa}=\frac{1}{\alpha}\left[\gamma+\psi\left(\frac{1}{2\kappa}\right)+\ln\left(2\beta\kappa\right)+\alpha\ln\left(m\right)+\kappa\right],
\label{eq:Equation30}
\end{equation}
where $\gamma=-\psi\left(1\right)$ is the Euler-Mascheroni constant and $\psi\left(z\right)=\Gamma^{'}\left(z\right)/\Gamma\left(z\right)$ is the digamma function, and the Theil \cite{Theil1967} index, $T_{\kappa}=\lim_{\theta\rightarrow1}GE_{\kappa}\left(\theta\right)$, defined as
\begin{equation}
\begin{split}
T_{\kappa}=&\frac{1}{\alpha}\left[\psi\left(1+\frac{1}{\alpha}\right)-\frac{1}{2}\psi\left(\frac{1}{2\kappa}-\frac{1}{2\alpha}\right)-\frac{1}{2}\psi\left(\frac{1}{2\kappa}+\frac{1}{2\alpha}\right)\right.\\
&\left.-\ln\left(2\beta\kappa\right)-\alpha\ln\left(m\right)-\frac{\alpha\kappa}{\alpha+\kappa}\right].
\end{split}
\label{eq:Equation31}
\end{equation}
Other GE indexes often used in applied work are the bottom-sensitive index,
\begin{equation}
GE_{\kappa}\left(-1\right)=-\frac{1}{2}+\frac{\Gamma\left(1+\frac{1}{\alpha}\right)\Gamma\left(1-\frac{1}{\alpha}\right)}{2\left[1+\left(\frac{\kappa}{\alpha}\right)^{2}\right]},
\label{eq:Equation32}
\end{equation}
and the top-sensitive index (or half the squared coefficient of variation),
\begin{equation}
GE_{\kappa}\left(2\right)=\frac{1}{2}CV^{2}_{\kappa}.
\label{eq:Equation33}
\end{equation}

Finally, the Atkinson index \cite{Atkinson1970} for inequality aversion parameter $\theta=1-\epsilon$ can be easily computed from $GE_{\kappa}\left(\theta\right)$ by exploiting the relationship
\begin{equation}
A_{\kappa}\left(\epsilon\right)=1-\left[\epsilon\left(\epsilon-1\right)GE_{\kappa}\left(1-\epsilon\right)+1\right]^{\frac{1}{1-\epsilon}},
\label{eq:Equation34}
\end{equation}
where $\epsilon\neq1$. The limiting form as $\epsilon\rightarrow1$ is
\begin{equation}
A_{\kappa}\left(1\right)=1-\exp\left(-MLD_{\kappa}\right).
\label{eq:Equation35}
\end{equation}

% Estimation ------------------------------------------------------------------------------------------------------------------------------------------

\subsection{Estimation}
\label{sec:Estimation}

Parameter estimation for the $\kappa$-generalized distribution can be performed using the Maximum Likelihood (ML) approach. Assuming that all observations $\mathbf{x}=\left\{x_{1},\ldots,x_{n}\right\}$ are independent, the likelihood function is
\begin{equation}
L\left(\boldsymbol{\theta};\mathbf{x}\right)=\prod\limits^{n}_{i=1}p\left(x_{i}\right)=\left(\alpha\beta\right)^{n}\prod\limits^{n}_{i=1}\frac{x_{i}^{\alpha-1}\exp_{\kappa}\left(-\beta x_{i}^{\alpha}\right)}{\sqrt{1+\beta^{2}\kappa^{2}x_{i}^{2\alpha}}},
\label{eq:Equation36}
\end{equation}
where $\boldsymbol{\theta}=\left\{\alpha,\beta,\kappa\right\}$ is the parameter vector. This leads to the problem of solving the partial derivatives of the log-likelihood function $l\left(\boldsymbol{\theta};\mathbf{x}\right)=\ln L\left(\boldsymbol{\theta};\mathbf{x}\right)$ with respect to $\alpha$, $\beta$ and $\kappa$. However, obtaining explicit expressions for the ML estimators of the three parameters is difficult, making direct analytical solutions intractable, and one needs to use numerical optimization methods.

Taking into account the meaning of the variable $x$, the mean value results to be equal to unity, i.e. $m=\int_{0}^{\infty}xp\left(x\right)\operatorname{d} x=1$. The latter relationship permits to express the parameter $\beta$ as a function of the parameters $\alpha$ and $\kappa$, obtaining
\begin{equation}
\beta=\frac{1}{2\kappa}\left[\frac{\Gamma\left(\frac{1}{\alpha}\right)}{\kappa+\alpha}\frac{\Gamma\left(\frac{1}{2\kappa}-\frac{1}{2\alpha}\right)}{\Gamma\left(\frac{1}{2\kappa}+\frac{1}{2\alpha}\right)}\right]^{\alpha}.
\label{eq:Equation37}
\end{equation}
In this way, the problem to determine the values of the free parameters $\left\{\alpha,\beta,\kappa\right\}$ of the theory from the empirical data reduces to a two parameter $\left\{\alpha,\kappa\right\}$ fitting problem. Therefore, to find the parameter values that give the most desirable fit, one can use the Constrained Maximum Likelihood (CML) estimation method \cite{Schoenberg1997}, which solves the general maximum log-likelihood problem of the form $l\left(\boldsymbol{\theta};\mathbf{x}\right)=\sum^{n}_{i=1}\ln p\left(x_{i};\boldsymbol{\theta}\right)^{w_{i}}$, where $n$ is the number of observations, $w_{i}$ the weight assigned to each observation, $p\left(x_{i};\boldsymbol{\theta}\right)$ the probability of $x_{i}$ given $\boldsymbol{\theta}$, subject to the non-linear equality constraint given by Equation \eqref{eq:Equation37} and bounds $\alpha,\beta>0$ and $\kappa\in\left[0,1\right)$. The CML procedure finds values for the parameters in $\boldsymbol{\theta}$ such that the negative of $l\left(\boldsymbol{\theta};\mathbf{x}\right)$ is minimized using the sequential quadratic programming method \cite{Han1977} as implemented, e.g., in \textsc{Matlab}\textsuperscript{\textregistered} 7.

% EMPIRICAL APPLICATION TO U.S. INCOME DATA -----------------------------------------------------------------------------------------------------------

\section{Empirical application to U.S. income data}
\label{sec:EmpiricalApplicationToUSIncomeData}

The $\kappa$-generalized distribution was fitted to data on personal income derived from the 2003 wave of the U.S. Panel Study of Income Dynamics (PSID) as released in the Cross-National Equivalent File (CNEF), a commercially available database compiled by researchers at Cornell University \cite{BurkhauserButricaDalyLillard2001}. The 2003 PSID-CNEF data have a sampling of 7,822 household, and all calculations are based on the household post-government income\textemdash i.e. the income recorded after taxes and government transfers\textemdash expressed in nominal local currency unit and normalized to its empirical average given by $31,812.39\pm598.74$ USD. We have omitted from the sample of incomes those with zero and negative value, and this affected only a tiny fraction of the data. Furthermore, incomes have been adjusted for differences in household size by dividing by the square root of the number of household members and weighted by the provided sampling weights \cite{Deaton1996}.

\begin{figure}[!hp]
\centering
\subfigure[Cumulative distribution]{\label{fig:Figure1_a}\includegraphics[width=0.45\textwidth]{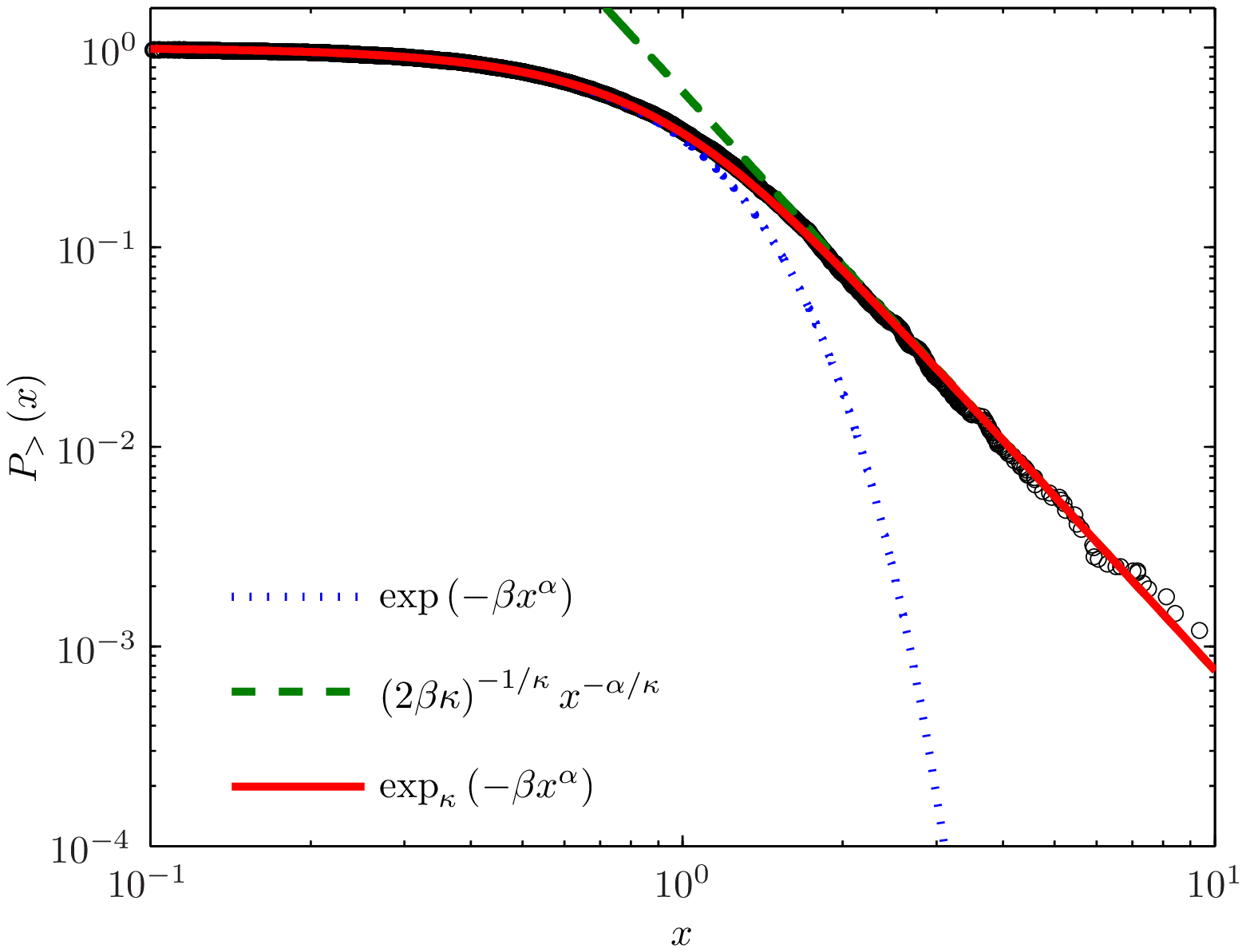}}
\subfigure[Density]{\label{fig:Figure1_b}\includegraphics[width=0.45\textwidth]{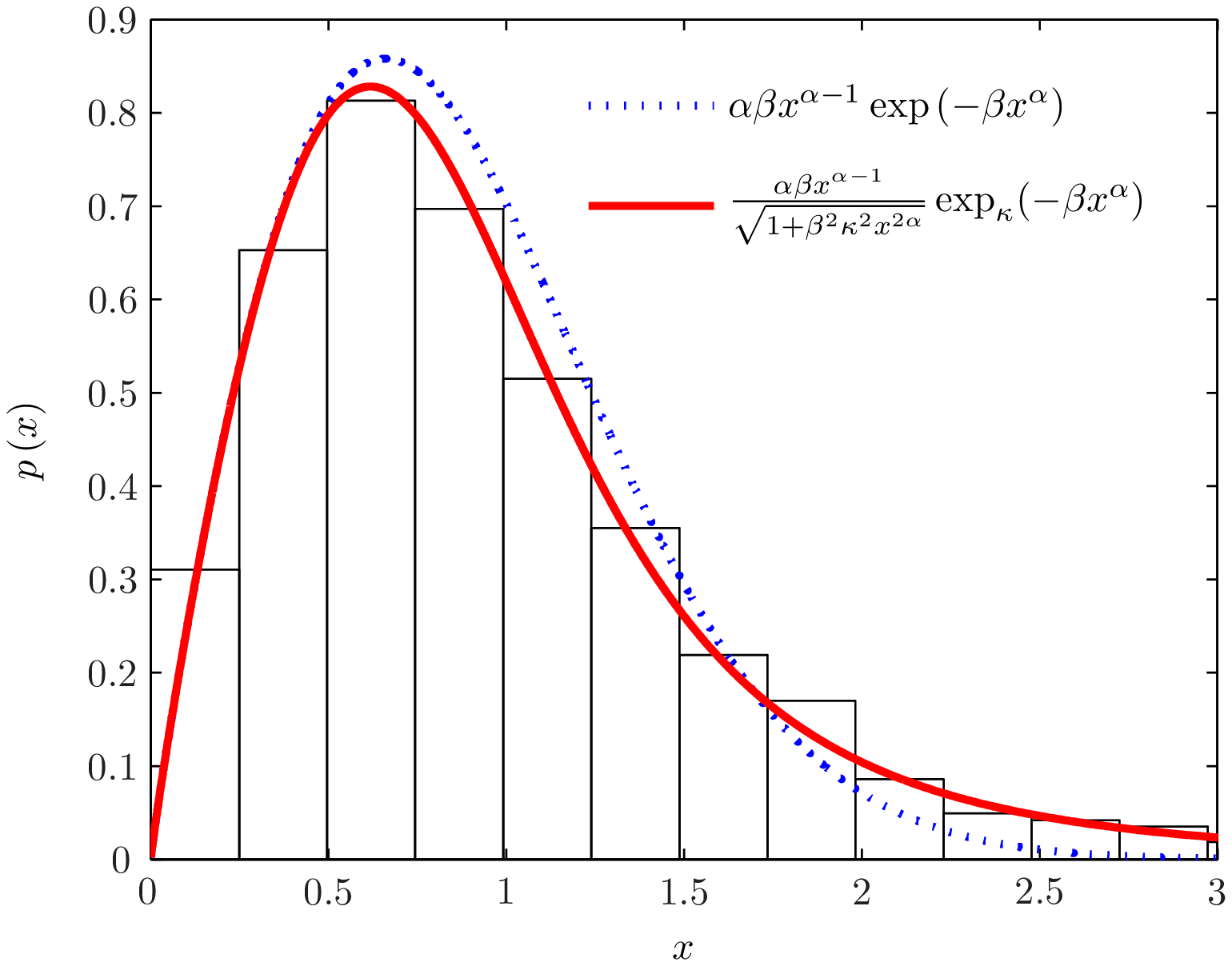}}
\subfigure[Lorenz curve]{\label{fig:Figure1_c}\includegraphics[width=0.45\textwidth]{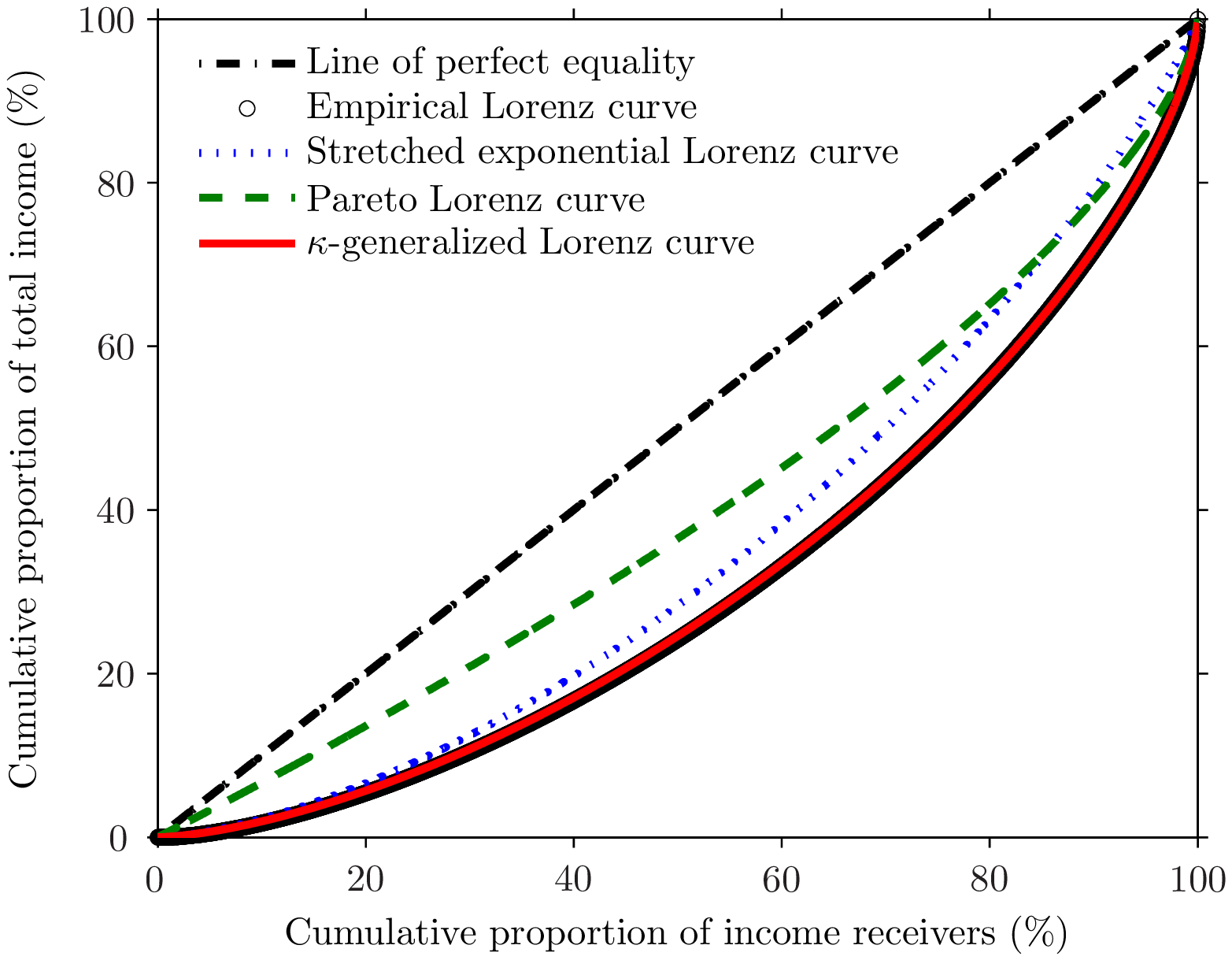}}
\subfigure[Q-Q plot]{\label{fig:Figure1_d}\includegraphics[width=0.45\textwidth]{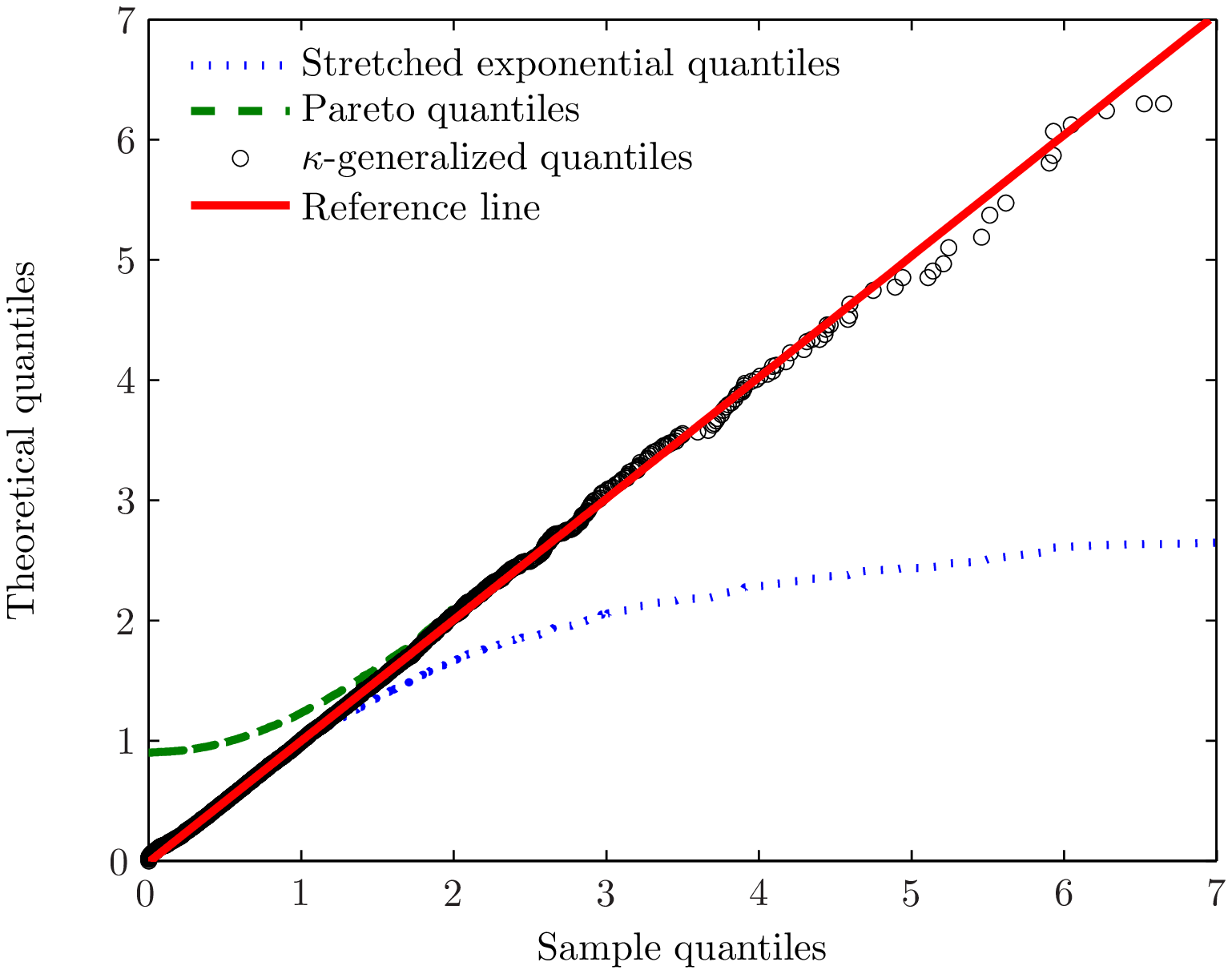}}
\caption{The mean-rescaled U.S. personal income distribution in 2003. \subref{fig:Figure1_a} Empirical cumulative distribution in the log-log scale. The solid line is our theoretical model given by Equation \eqref{eq:Equation11} fitting very well the data in the whole range from the low to the high incomes including the intermediate income region. This function is compared with the ordinary stretched exponential one (dotted line)\textemdash fitting the low income data\textemdash and with the pure power-law (dashed line)\textemdash fitting the high income data. \subref{fig:Figure1_b} Probability density histogram with superimposed fits of the $\kappa$-generalized (solid line) and Weibull (dotted line) densities. \subref{fig:Figure1_c} Lorenz curve. The hollow circles represent the empirical data points and the solid line is the theoretical curve given by Equation \eqref{eq:Equation27} using the same parameter values as in panels \subref{fig:Figure1_a} and \subref{fig:Figure1_b}. The dash-dot line corresponds to the Lorenz curve of a society in which everybody receives the same income and thus serves as a benchmark case against which actual income distribution may be measured. The dotted and dashed lines represent the theoretical Lorenz curves from the stretched exponential and Pareto distributions given by Equations Equation \eqref{eq:Equation38a} and Equation \eqref{eq:Equation38b}, respectively. \subref{fig:Figure1_d} Q-Q plot of the sample quantiles versus the corresponding quantiles of the fitted $\kappa$-generalized (hollow circles), stretched exponential (dotted line) and Pareto (dashed line) distributions. Where not displayed, the quantiles of these last two distributions coincide with those of the $\kappa$-generalized. The reference (solid) line has been obtained by locating points on the plot corresponding to around the 25\textsuperscript{th} and 75\textsuperscript{th} percentiles and connecting these two. In panels \subref{fig:Figure1_a}, \subref{fig:Figure1_b} and  \subref{fig:Figure1_d} the income axis limits have been adjusted according to the range of data to shed light on the intermediate region between the bulk and the upper end of the distribution.}
\label{fig:Figure1}
\end{figure}

The best-fitting parameter values were determined using CML estimation as discussed in Section \ref{sec:Estimation}. This resulted in the following estimates: $\alpha=1.9115\pm0.0003$, $\beta=1.0568\pm0.0002$ and $\kappa=0.6587\pm0.0003$. The very small value of the errors indicates that the parameters were precisely estimated, and the comparison between the observed and fitted probabilities in panels \subref{fig:Figure1_a} and \subref{fig:Figure1_b} of Figure \ref{fig:Figure1} suggests that the $\kappa$-generalized distribution offers a great potential for describing the data over their whole range, from the low to medium income region through to the high income Pareto power-law regime, including the intermediate region for
which a clear deviation exists when two different curves are used.
%
% -----------------------------------------------------------------------------------------------------------------------------------------------------
% -----------------------------------------------------------------------------------------------------------------------------------------------------

Panel \subref{fig:Figure1_c} of the same figure depicts the data points for the empirical Lorenz curve, i.e. $L\left(i/n\right)=\sum^{i}_{j=1}x_{j}/\sum^{n}_{j=1}x_{j}$, $i=1,2,\ldots,n$, superimposed by the theoretical curve $L_{\kappa}\left(u\right)$ given by Equation \eqref{eq:Equation27} with estimates replacing $\alpha$ and $\kappa$ as necessary. This formula is shown by the solid line in the plot, and fits the data exceptionally well. The plot also compares the empirical Lorenz curve to the theoretical ones associated with the stretched exponential and Pareto distributions, respectively given by
\begin{subequations}
\begin{equation}
\lim_{\kappa\rightarrow0}L_{\kappa}\left(u\right)=P\left(1+\frac{1}{\alpha},-\ln\left(1-u\right)\right),
\label{eq:Equation38a}
\end{equation}
where $P\left(\cdot,\cdot\right)$ is the lower regularized incomplete gamma function, and
\begin{equation}
\lim_{x\rightarrow\infty}L_{\kappa}\left(u\right)=1-\left(1-u\right)^{1-\frac{\kappa}{\alpha}}.
\label{eq:Equation38b}
\end{equation}
\end{subequations}
As one can easily recognize, these curves account for only a small part of the whole story.

In order to provide indirect checks on the validity of the parameter estimation, we have also calculated the sample values of the Gini and Theil indexes, obtained respectively as $G=n^{-2}\sum^{n}_{i=1}\left(2i-n-1\right)x_{i}$ and $T=n^{-1}\sum^{n}_{i=1}x_{i}\ln\left(x_{i}\right)$, which return $G=0.3805\pm0.0092$ and $T=0.2790\pm0.0295$. The corresponding predictions from the analytical expressions Equation \eqref{eq:Equation28} and Equation \eqref{eq:Equation31} are $G_{\kappa}=0.3780$ and $T_{\kappa}=0.2600$, and result completely covered by the 95\% confidence intervals constructed around the empirical values.\footnote{The confidence intervals for the observed Gini and Theil indexes have been calculated via the bootstrap resampling method based on 1000 replications \cite{MillsZandvakili1997}.}

The accuracy of our distributional model was further examined by testing the hypothesis that the observed data follow a $\kappa$-generalized distribution through the Kolmogorov-Smirnov (K-S) goodness-of-fit test statistic given by $D^{+}=\max_{1\leq i\leq n}\left[in^{-1}-P_{<}\left(x_{i}\right)\right]$, $i=1,2,\ldots,n$. Since in this case there is no asymptotic formula for calculating the $p$-value, we have reduced the problem to testing that the $x$ values have a standard exponential distribution (i.e., an exponential distribution with parameter equal to 1) by relating the function $P_{\geq}\left(x\right)$ given by Equation \eqref{eq:Equation11} to the ordinary exponential function, namely $\exp_{\kappa}\left(-\beta x^{\alpha}\right)=\exp\left(-x_{\kappa}\right)$, through the transformation $x_{\kappa}=\kappa^{-1}\log\left(\sqrt{1+\beta^{2}\kappa^{2}x^{2\alpha}}+\beta\kappa x^{\alpha}\right)$, where the parameters are estimated from the data. Thus the significance level in the upper tail is given approximatively by $P_{\geq}\left(T^{\ast}\right)=\exp\left[-2\left(T^{\ast}\right)^{2}\right]$, with $T^{\ast}=D^{+}\left(\sqrt{n}+0.12+0.11/\sqrt{n}\right)$ \cite{Stephens1970}. The results are $D^{+}=0.0085$ and $P_{\geq}\left(T^{\ast}\right)=0.3263$, and state that the maximum distance between the empirical data and the theoretical model as assessed by the K-S statistic is so small that the $p$-value is not able to lead to rejection of the null hypothesis that the data may come from a $\kappa$-generalized distribution at any of the usual significance levels (1\%, 5\% and 10\%). The linear behavior emerging from the Quantile-Quantile (Q-Q) plot of the sample quantiles versus the corresponding quantiles of the fitted $\kappa$-generalized distribution and its two limiting cases displayed in panel \subref{fig:Figure1_d} of Figure \ref{fig:Figure1} confirms the quantitative results obtained by hypothesis testing, as well as the fact that the stretched exponential and Pareto distributions can give only a partial and incomplete description of the data.

% CONCLUDING REMARKS ----------------------------------------------------------------------------------------------------------------------------------

\section{Concluding remarks}
\label{sec:ConcludingRemarks}

Fitting a parametric model to income data can be a valuable and informative tool for distributional analysis. Not only can one summarize the information contained in thousands of observations, but also useful information can be drawn directly from the estimated parameters. For example one could be interested in measuring income inequality, comparing different distributions or elaborating income redistribution policy: these concepts may be directly derived from parameters of a fitted distribution.

Starting from the Pareto contribution, a wide variety of functional forms have been considered as possible models for the distribution of personal income by size, and other approaches can no doubt be suggested and deserve attention.

In this work we have proposed a new fitting function having its roots in the framework of the $\kappa$-generalized statistical mechanics. The model has a bulk very close to the stretched exponential one\textemdash which is recovered when the deformation parameter $\kappa$ tends to zero\textemdash while for high values of income its upper tail approaches a Pareto distribution, thus being able to describe the data over the entire range. The performance of the distribution has been checked against real data on personal income for the United States in 2003 and has been found to fit remarkably well. The analysis of inequality performed in terms of its parameters reveals the merit of the new proposed distribution, and provides the basis for a fruitful interaction between the two fields of statistical mechanics and economics.

% REFERENCES ------------------------------------------------------------------------------------------------------------------------------------------

% EOF -------------------------------------------------------------------------------------------------------------------------------------------------

\end{document}